\documentclass{article}
\usepackage{amsmath}
\usepackage{amssymb}
\usepackage{amsfonts}
\usepackage[backend=bibtex,url=false,doi=false,style=numeric-comp,firstinits=true]{biblatex}
\bibliography{/home/rsrsl/library/jabref_biblio,biblio}
\usepackage{graphicx}
\usepackage{float}
\usepackage{textcomp}
\usepackage{booktabs}
\usepackage{algorithm}
\usepackage{xcolor}

\title{A~benchmark~for~data-based~office~modeling:\\ challenges related to CO$_2$ dynamics}
 \author{Riccardo S. Risuleo, Marco Molinari, Giulio Bottegal,\\ H\r{a}kan
 Hjalmarsson, and Karl H. Johansson
\thanks{Riccardo S. Risuleo, Marco Molinari, Giulio Bottegal, H\r{a}kan
  Hjalmarsson, and Karl H. Johansson are with the ACCESS Linnaeus Center, School
  of Electrical Engineering, KTH Royal Institute of Technology, Stockholm,
Sweden (e-mail: \{risuleo; marcomo; bottegal; hjalmars; kallej\}@kth.se)\}
  The research leading to these results has received funding from the European
  Union Seventh Framework Programme [FP7/2007-2013] 
  under grant agreement n\textdegree{} 257462 HYCON2 Network of excellence, the
  European Institute of Technology (EIT) Information and Communication
  Technology (ICT) Labs, the Swedish Energy Agency, the Swedish Governmental
  Agency for Innovation Systems (VINNOVA), the Knut and Alice Wallenberg
  Foundation, the European Research Council under the advanced grant LEARN,
  contract 267381 and from the Swedish Research Council under contract
  621-2009-4017.}} 

\begin{document}
\maketitle

\begin{abstract}
This paper describes a benchmark consisting of a set of synthetic measurements
relative to an office environment simulated with the software \textsc{IDA-ICE}.
The simulated environment reproduces a laboratory at the KTH--EES Smart 
Building, equipped with a building management system.  The data set contains
measurement records collected over a period of several days. The signals
correspond to CO$_2$ concentration, mechanical ventilation
airflows, air infiltrations and occupancy. Information on door and window opening
is also available.  This benchmark is intended for testing data-based modeling
techniques. The ultimate goal is the development of models to improve the forecast
and control of environmental variables.  Among the numerous challenges related
to this framework, we focus on the problem of occupancy estimation using
information on CO$_2$ concentration, which we treat as a blind identification
problem. For benchmarking purposes, we present two different identification
approaches: a baseline overparameterization method and a kernel-based method.
\end{abstract}


\section{Introduction}
The recent development of advanced control and monitoring techniques in
buildings has shown promising results for the reduction of energy use and the
improvement of indoor comfort (see
e.g.~\cite{vana2014model},~\cite{siroky2011experimental},~\cite{ferreira2012neural}
and~\cite{costa2013building}). Two key factors that
affect the quality of the indoor environment are temperature and CO$_2$
concentration. Many efforts have consequently been devoted to developing novel smart
and energy-efficient control strategies to guarantee human comfort by acting
over such environmental variables. Among the several possible control
techniques, a promising direction seems to be the deployment of model predictive
control (MPC) (see~\cite{oldewurtel2012use} and~\cite{parisio2014implementation}).
Model-based control performance benefits from models able to capture
accurately the system dynamics. These models are often derived using first-order
principles. However, this approach might not always be possible, due
to incomplete knowledge of the building characteristics, model complexity
issues, unpredictable dynamics and cost constraints. In these cases it
is interesting to explore the potential given by automatic data-based
modeling techniques.

Motivated by these aspects, in this paper we present a set of simulated data
regarding actuation signals and environmental variables affecting the comfort
conditions of a specific office room. The data are generated using
IDA-ICE~\citeauthor{equa2015equa}, a well-established simulator for building
dynamics~\cite{crawley2008contrasting}. The data set includes the
variables that mostly influence the CO$_2$ concentration, namely
ventilation, number of occupants and infiltrations through doors and windows. The
simulations span a period of one week and involve different environmental
conditions, such as low/high number of occupants and window and door opening.
The simulated environment models a laboratory used at the School of Electrical
Engineering at KTH\@. The use of simulations, compared to measurements, enables a
more refined control over the experimental conditions while still capturing the
main dynamics of the system. The rationale behind this data set is to,
\begin{enumerate}
  \item Assess the capacity of system identification techniques of successfully
    capturing the dynamics of the simulated environment;
  \item Offer a benchmark on
    which the current state-of-the-art system identification algorithms can be
    compared.
\end{enumerate}

An additional contribution of this paper is the discussion of some of the
(many) possible system identification challenges arising when dealing with
CO$_2$ dynamics. Among these challenges, we focus on modeling the dynamic
relation between occupancy of the room and CO$_2$ concentration.
In fact, occupancy affects the indoor environment through heat gains and
CO$_2$. Occupancy estimation is crucial to determine the evolution of indoor
environmental conditions. The problem of occupancy estimation has been
addressed in literature in several ways (see
e.g.~\cite{han2012occupancy},~\cite{liao2012agent},~\cite{ai2014occupancy},~\cite{ebadat2013estimation}).
Here we tackle this problem by proposing occupancy estimation from
CO$_2$ measurements. Assuming that no data records on the occupants are
available, we cast this problem as a \emph{blind system identification
problem}~\cite{abedmeraim1997blind}. We describe and
test two algorithms: One of them is based on overparameterization and is
inspired by~\cite{bai1998optimal}; the other is described in~\cite{bottegal2015blind}.

The paper is organized as follows. In Section 2, we introduce the
physical characteristics of the simulated environment. In Section~\ref{sec:dataset}, we describe the data set generated with the simulator. In
Section~\ref{sec:schematic}, we provide some insights on the dynamics of
CO$_2$. In Section~\ref{sec:blind_id}, we propose a challenge based on blind
identification of the number of occupants. Some conclusions end the paper.

\section{The physical and simulated environments}
\subsection{Motivations}
The work presented in this paper was carried out within the research activities
at the KTH-EES Smart Building. The building, located in the KTH main campus in
Stockholm, hosts offices and laboratories and is equipped with indoor and
outdoor environmental sensors. Currently, two rooms of the buildings are used
for experimental testing of advanced controls schemes. One of the rooms was
chosen for the simulation as the physical characteristics of the KTH-EES Smart
Building make it a good representative of office buildings in Sweden. In
addition, the availability of sensors and actuators allow us to validate the
simulations against real data.


\subsection{Geometry description of the room}

\begin{figure}[htb]
\centering
\includegraphics[width=50mm]{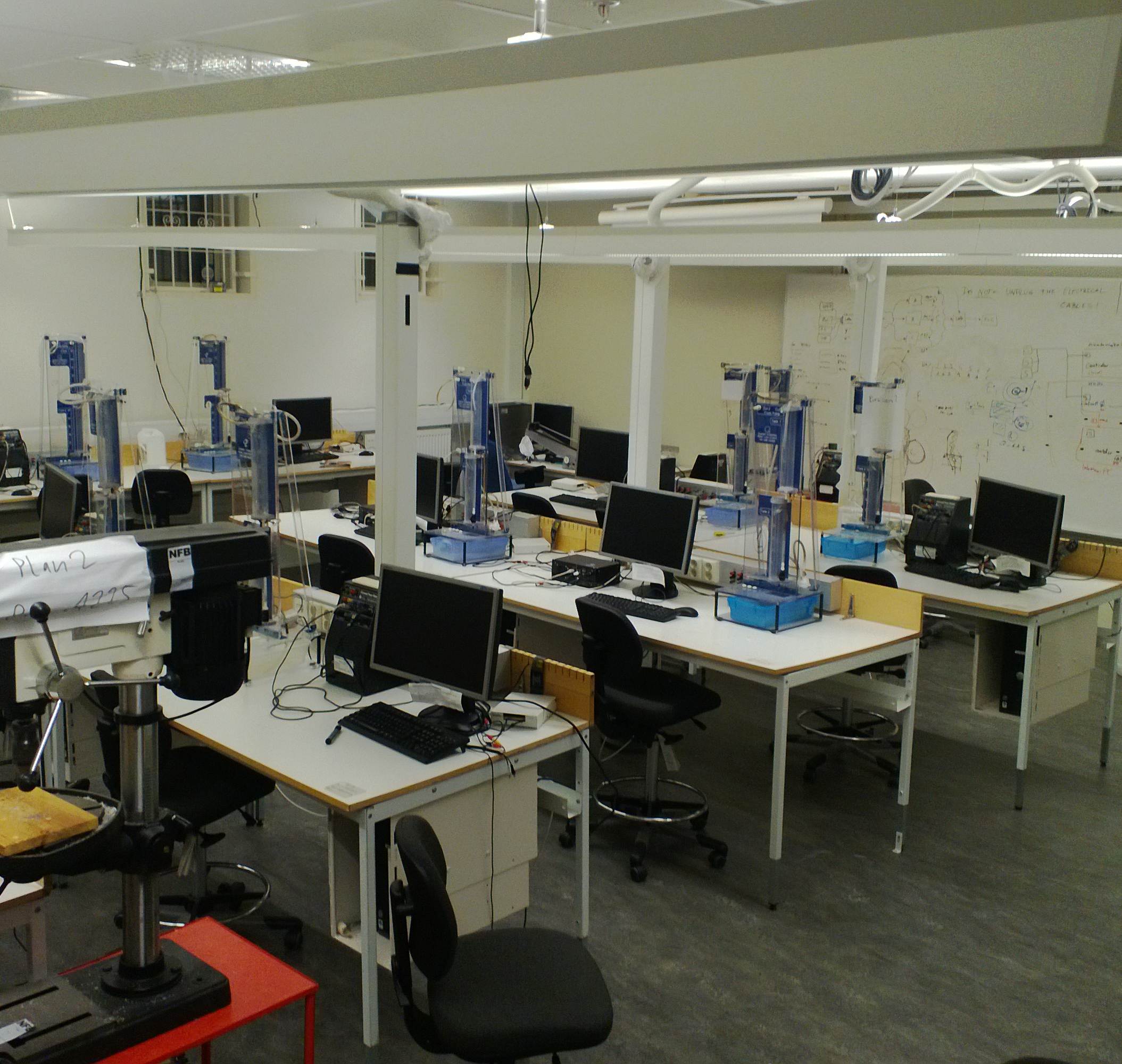}
\caption{A picture of the laboratory at KTH simulated for the benchmark.}
\label{fig:testebd_pic}
\end{figure}

The model simulated in this paper represents a laboratory room of 80 m$^2$
footprint (Fig.~\ref{fig:testebd_pic}); the room has four small external
windows with a total area of approximately 2.5 m$^2$. The laboratory is used
for lecturing groups of students; the occupancy level is hence rather variable,
ranging from periods of no occupancy to peaks of more than 20 students.

Mechanical ventilation in the room is provided between 8:00 and 18:00 with a variable rate ventilation system, with
ventilation air flows ranging from 0.08 m$^3$/s to 0.28 m$^3$/s.
The ventilation air flow is determined by the CO$_2$ concentration in the room.

\subsection{Simulation software environment}

The generation of CO$_2$ data was carried out via IDA-ICE 4.6. IDA-ICE is a
commercial program for dynamic simulations of energy and comfort in
buildings; it features equation-based modeling (NMF-language or Modelica
language~\cite{fritzson2010principles}) and is equipped with a variable timestep
differential-algebraic (DAE) solver~\cite{sahlin2004whole}.

\subsection{Validation of the generated data}

\begin{figure}[htb]
\centering
\includegraphics[trim = 3.4cm 8.5cm 3.5cm 8.5cm,clip,width=1\columnwidth]{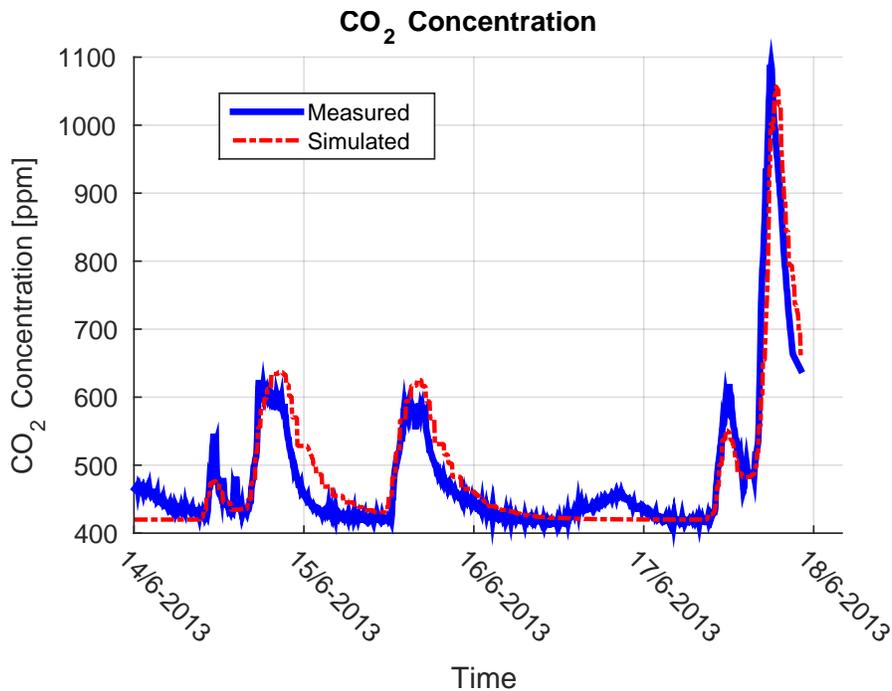}
\caption{Validation of the IDA-ICE model. CO$_2$ levels from room measurements
and from simulation are compared.}
\label{fig:Physical_model_validation}
\end{figure}

In order to test the accuracy of the IDA-ICE physical model with respect to the
real room dynamics, simulated and measured data for CO$_2$ were compared in
Fig.~\ref{fig:Physical_model_validation}, under the same conditions of
occupancy, ventilation and window opening. The two sets of measured and
simulated data show that the physical model is capable of capturing the main
CO$_2$ dynamics within the room space. The mismatch between the two curves is
attributed to events whose effect, though minor, is not simple to account for;
examples of such events are doors kept open and undetected window openings.

\section{Description of the data set}\label{sec:dataset}
\begin{figure}[htb]
\centering
\includegraphics[width=.98\columnwidth]{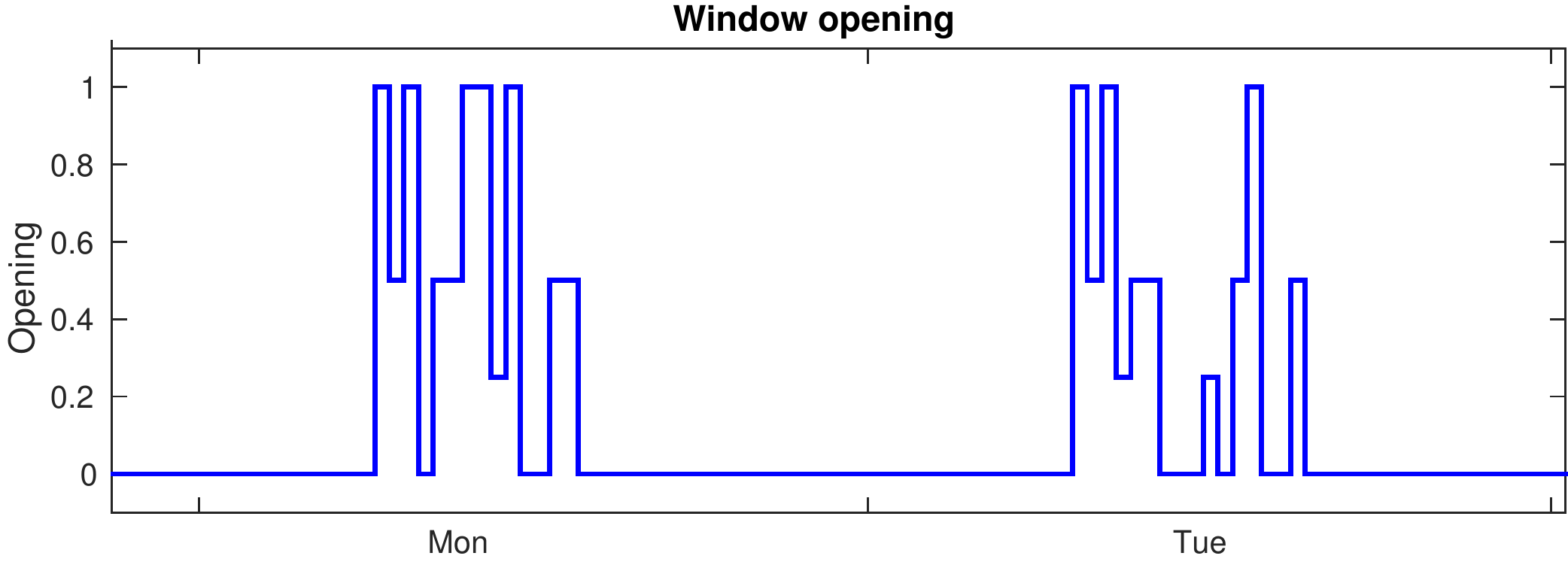}\\
\includegraphics[width=.98\columnwidth]{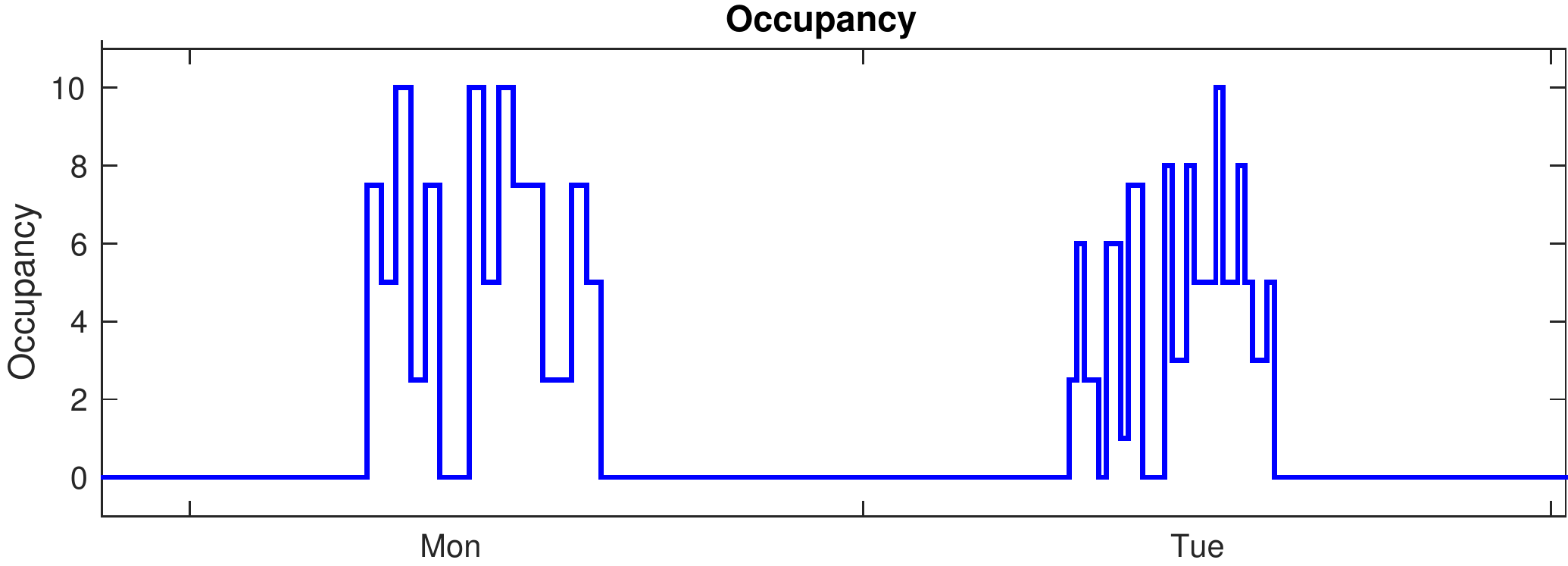}
  \caption{Examples of window opening and occupancy conditions. Profiles for the first two days of the simulated week are shown. Top: window
opening signal. Bottom:
number of occupants in the office.}\label{fig:occupancy_pattern}
\end{figure}

The data simulate the office environment during a summer week, from July 13th,
2014 to July 19th, 2014. The climatic conditions are relative to Stockholm;
they were collected at the Bromma Airport meteorological station and issued by
the Swedish Meteorological and Hydrological Institute. The input variables for
the CO$_2$ data generation are zone occupancy, ventilation and air
infiltrations. We simulate different conditions of occupancy; an example is
shown in
Fig.~\ref{fig:occupancy_pattern}. We denote such conditions as low/medium/high
level occupancy. The rationale behind this is that, due to the nonlinearity of
the control system, the system dynamics can change over the different
occupancy levels. This will be explained in the next section.

In the IDA-ICE model, CO$_2$ generation is a function of the activity of the
occupants. In these simulations, the activity levels are set to 1.8 Metabolic
Equivalent of Task (MET), corresponding to a light physical activity, which
resembles typical office working conditions.

The building air tightness is assumed to be 1.5 Air Changes per Hour (ACH) at
50 Pa, corresponding to a standard building in Sweden. Air infiltrations
are allowed through doors and windows depending on the wind speed.
We simulate two different conditions, related to occupants' behavior:
\begin{enumerate}
\item Windows are kept closed for the whole time span;
\item Windows are opened at varying percentages.
\end{enumerate}
An example of the second situation is depicted in
Fig.~\ref{fig:occupancy_pattern}, which shows the percentage of one window
opening as function of time.

The different conditions on windows opening and occupancy level are combined
together, giving rise to 6 data sets. The simulation outcomes are collected in
files in the Matlab workspace \texttt{.mat} format. They can be downloaded both
from the KTH EES Smart Building project web page (see~\citeauthor{kth-ees2015ees}
in the reference list), and the IFAC TC 1.1 Repository database (see
~\citeauthor{ifac2015ifac}). Features of the data sets, together with
relative file names, are summarized in Table~\ref{tab:datasets}.

\begin{table}[h!]\label{tab:datasets}
\begin{center}
\begin{tabular}{lcc}
File name & Occupancy level & Windows \\
\toprule
\texttt{kth\_lowc.mat} & low	& closed \\
\texttt{kth\_mowc.mat} & medium	& closed \\
\texttt{kth\_howc.mat} & high	& closed \\
\texttt{kth\_lowo.mat} & low	& open \\
\texttt{kth\_mowo.mat} & medium	& open \\
\texttt{kth\_howo.mat} & high	& open \\
\bottomrule
\end{tabular}
\vspace{2mm}
\caption{Features of the simulated data sets.}
\end{center}
\end{table}

Each \texttt{.mat} file consists of a number of vectors collecting the samples
of the simulated variables. They are listed below.
\begin{itemize}
\item \texttt{occupancy}: number of people;
\item \texttt{CO2}: noiseless CO$_2$ concentration;
\item \texttt{CO2\_noise}: CO$_2$ concentration with additive Gaussian measurement noise;
\item \texttt{outflow\_leakages}: overall air outflow due to infiltrations;
\item \texttt{inflow\_leakages}: overall air inflow due to infiltrations;
\item \texttt{outflow\_ventilation}: air outflow due to ventilation;
\item \texttt{inflow\_ventilation}: air inflow due to ventilation;
\item \texttt{ventilation\_control}: ventilation control signal;
\item \texttt{window\_opening}: window opening percentage.
\end{itemize}

Fig.~\ref{fig:Physical_model_validation} shows that there
is a considerable amount of noise in the measurements. To a generate realistic
dataset, we added Gaussian white noise to the output of IDA-ICE, which is
noiseless. The covariance of the added noise was tuned to obtain a
signal-to-noise ratio of 10 dB, which agrees with the noise covariance
estimated from the data in Fig.~\ref{fig:Physical_model_validation}.

The maximum integration time step is set to three minutes, which means that the
IDA-ICE internal solver is forced to provide the integral solution at a maximum
three minutes interval, even if the program is still allowed to choose shorter
time steps. The output time step for the solutions is also set to three minutes;
hence, the vectors in the \texttt{.mat} file contain 3360 entries.

\section{Description of room dynamics and control architecture}\label{sec:schematic}
\begin{figure}[htb]
\centering
\includegraphics[width=0.7\columnwidth]{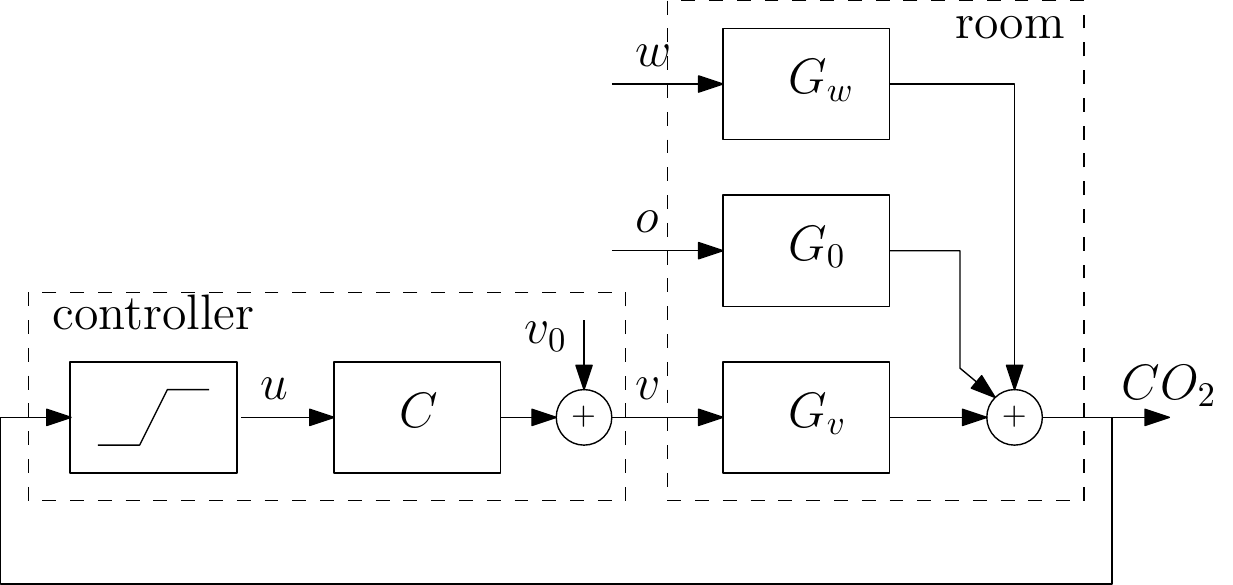}
\caption{Block scheme of room dynamics and control architecture. The CO$_2$ dynamics are driven by the ventilation $v$, the occupancy $o$ and window opening $w$. The measured CO$_2$ is the input to a controller constituted by two blocks in cascade. The first one is a static saturation, the second one is a linear filter denoted by $C$.}
\label{fig:block_scheme}
\end{figure}
A schematic representation of the whole dynamics of interest is depicted in
Fig.~\ref{fig:block_scheme}. The signal CO$_2(t)$ can be thought of as the sum
of three contributions. The first is given by possibly open windows, which are
represented by the signal $w(t)$ and influence the output through the system
$G_w$. The second contribution is given by the occupancy, denoted by $o(t)$,
and the related dynamic system $G_o$. The third one is the result of the air
ventilation acting on the room. The ventilation, denoted by $v(t)$, is driven
by a specific control system. The controller can be seen as the cascade of a
static nonlinearity, which acts as a saturation, and a linear controller, plus
a constant source signal $v_0$. The saturation receives the current value of
the CO$_2$ concentration and transforms it into the signal $u$ according to the
following map:
\begin{equation}\label{eq:saturator}
u(t) = \left\{ \begin{array}{lcl}
				0 & \mbox{if} & \textrm{CO}_2(t) < 700\,\textrm{ppm},\\
				\frac{\textrm{CO}_2(t)-700}{1100-700} & \mbox{if} &  700\,\textrm{ppm} \leq\,\textrm{CO}_2(t) \leq 1100\,\textrm{ppm},\\
				1 & \mbox{if} & \textrm{CO}_2(t) > 1100\,\textrm{ppm}.  \end{array} \right.
\end{equation}
Note that $u(t)$ is not available to the experimenter. This signal is filtered
by a linear filter (denoted by $C$ in Fig.~\ref{fig:block_scheme}), which is a
PID controller with unknown parameters. The resulting signal is then summed to
a constant value $v_0$, which provides a constant base ventilation to the room.
\subsection{Related system identification problems}
The dynamics described above give rise to several problems related to unmodeled
dynamics. Knowing the models $G_w$, $G_o$, $G_v$ and the controller
architecture is a basic requirement to design intelligent regulation
strategies. Quite unfortunately (or perhaps, from a system identification
perspective, \emph{luckily}), getting the aforementioned models seems to be a
challenging task. This mainly because of three reasons:
\begin{enumerate}
\item Although the room dynamics could in principle be quite-well approximated by linear systems, these systems could be time-varying, due to seasonal phenomena, etc.;
\item There is a number of non modelable phenomena (air leakages, computers, etc.) which might influence the room dynamics and should be regarded as noise;
\item Some signals, such as $w(t)$ and $o(t)$ in
  Fig.~\ref{fig:block_scheme}, may not be available in practice.
\end{enumerate}
We point out some of the possible system identification problems arising from
the proposed benchmark.
\begin{itemize}
\item \emph{Identification of the overall room dynamics.} Perhaps this is the
  most important task; as mentioned above, estimating models for $G_w$, $G_o$,
  $G_v$, is a basic requirement for designing advanced control strategies.
  Assuming all the inputs are available, this is a closed-loop identification
  problem, where several disturbances are acting on the system.
\item \emph{Identification of the controller}. The knowledge of the control
  algorithm is important in applications such as diagnostics and fine tuning of
  the regulation system. Assuming that the saturation is unknown, this problem
  can be seen as a Hammerstein system identification problem, with the added
  complication of the closed-loop.
\item \emph{Identification of dynamic relation between occupancy and CO$_2$}.
  Since occupants have high impact on the CO$_2$, this constitutes an
  interesting problem. Assuming that no information about $o(t)$ is available,
  this problem is a blind system identification problem, with the unknown input
  signal being piecewise constant. The problem can be relaxed by assuming that
  knowledge about door opening is available, which determines the time instants
  at which the input might change value. In the next section, we propose two
  algorithms for this problem and we test them on our benchmark data.
\end{itemize}

\section{The blind system identification challenge}\label{sec:blind_id}
In this section, we address the problem of identifying the dynamic relation between occupancy and CO$_2$. We assume we do not have access to the occupancy signal $o$, which is piecewise constant, but, having installed a sensor on the door of the room, we know when this signal may change value.

With reference to Fig.~\ref{fig:block_scheme}, the dynamics of CO$_2$ as function of $o$ can be described by the following closed-loop transfer function
\begin{equation}
Q = \frac{G_o}{1-C G_v} \,,
\end{equation}
where we have neglected the presence of the saturation. We consider a linear time-invariant model for $Q$. Furthermore, we define the new output $\overline{\textrm{CO}}_2(t) : = \textrm{CO}_2(t) - \textrm{CO}_{2,0}$, where $\textrm{CO}_{2,0}$ is the outdoor CO$_2$ concentration (see Table 3 in Appendix) and model general uncertainties as white noise. Then we can rewrite the model in time domain
\begin{equation}\label{eq:id_fir}
\overline{\textrm{CO}}_2(t) = \sum_{k=1}^n q(k) o(t-k) + e(t) \,,
\end{equation}
where we have approximated the system dynamics with a (long) FIR of order $n$.
The term $e(t)$ is the prediction error that contains the
measurement noise and is modeled as Gaussian white noise.
We collect $N$ samples of the output. Introducing a vector notation, we rewrite~\eqref{eq:id_fir} as a linear regression problem, i.e.
$\overline{\textrm{CO}}_2 = O q + e$, where $O$ is a suitable Toeplitz matrix
containing $o$.
If we denote the door opening events by $T_1,T_2\dots T_p=N$ and define the matrix $ H = \mathrm{diag}\{\mathbf{1}_{T_1},\mathbf{1}_{T_2-T_1}\dots\mathbf{1}_{T_p-T_{p-1}}\}$, then we can write
$o = Hx$, where $x \in \mathbb{R}^p$ denotes the unknown occupancy levels. Using this notation, we now give two algorithms for this problem.
\subsection{Benchmarking algorithms}
\subsubsection{Baseline method}
This method is a re-adaptation of the Hammerstein system identification method
proposed in~\cite{bai1998optimal}. It consists of the following steps.
\begin{enumerate}
\item Define $\Phi = \begin{bmatrix}H & SH & S^2H & \dots
  &S^{p-1}H\end{bmatrix}$, where  $S$ acts as one-position downwards shifting matrix, so that we can rewrite $\overline{\textrm{CO}}_2 = \Phi \theta  + e$, where $\theta := \mathrm{vec}(qx^T)$.
\item Compute a least-squares estimate of $\theta$ and denote it by $\hat \theta$.
\item Form the $n \times p$ matrix $\hat \Theta$ by reshaping $\hat \theta$.
\item Compute $\hat q$ as the first left singular vector of $\hat \Theta$ and $\hat x$ as the first right singular vector of $\hat \Theta$.
\end{enumerate}
A nice property of this method is that it can be proven to be asymptotically
consistent~\cite{bai1998optimal}.

\subsubsection{Kernel-based method}
We test a recently proposed blind system identification method tailored for
this type of problem. It is based on kernel-based methods combined with the
so-called \emph{stable spline kernel}~\cite{pillonetto2014kernel}. Due to
space constraints, we do not provide details on this method here, referring the
interested reader to~\cite{bottegal2015blind}.

\subsection{Testing the benchmarking algorithms on data}
\begin{figure*}[htb]\label{fig:co2_test}
\centering
\includegraphics[width=.98\columnwidth]{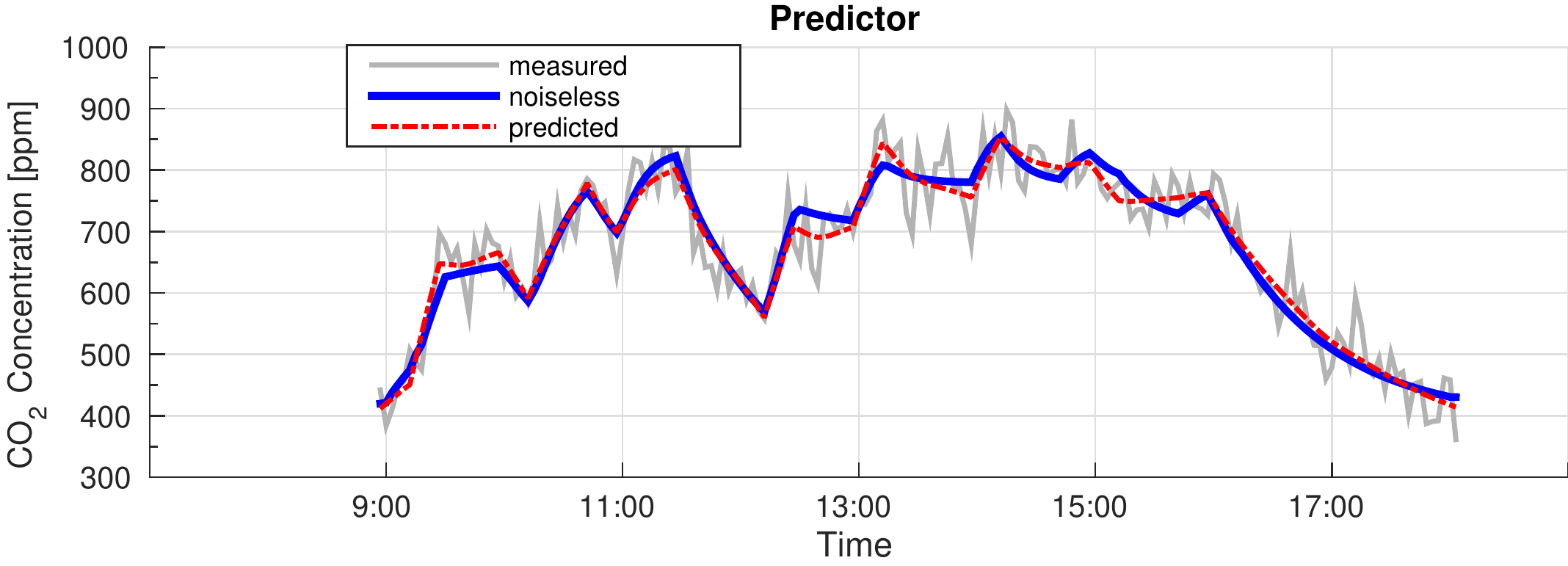}\\
\includegraphics[width=.98\columnwidth]{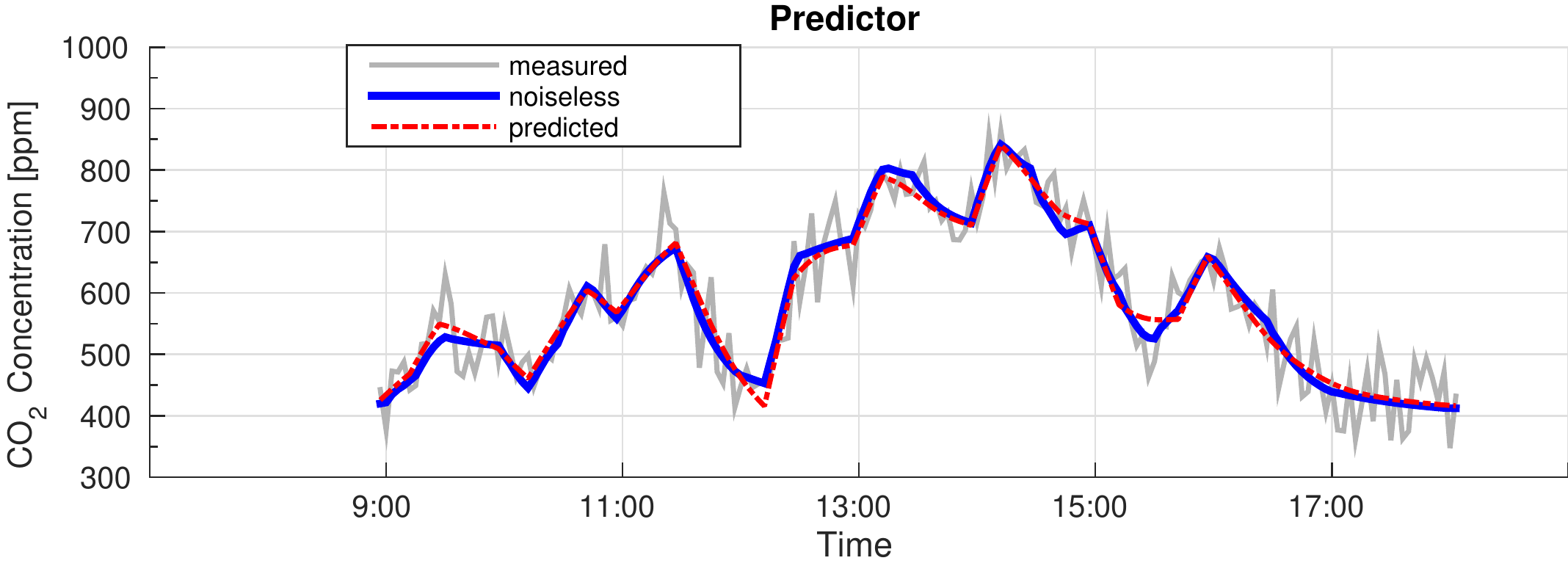}\\
\includegraphics[width=.98\columnwidth]{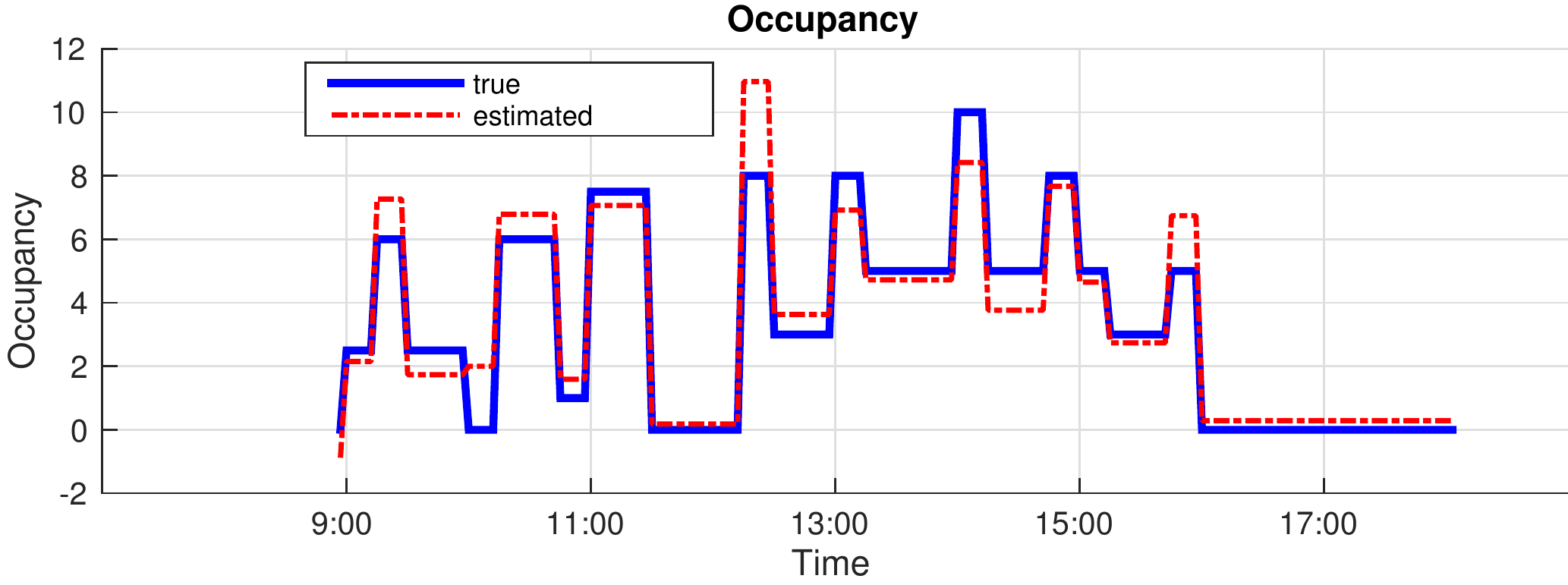}\\
\includegraphics[width=.98\columnwidth]{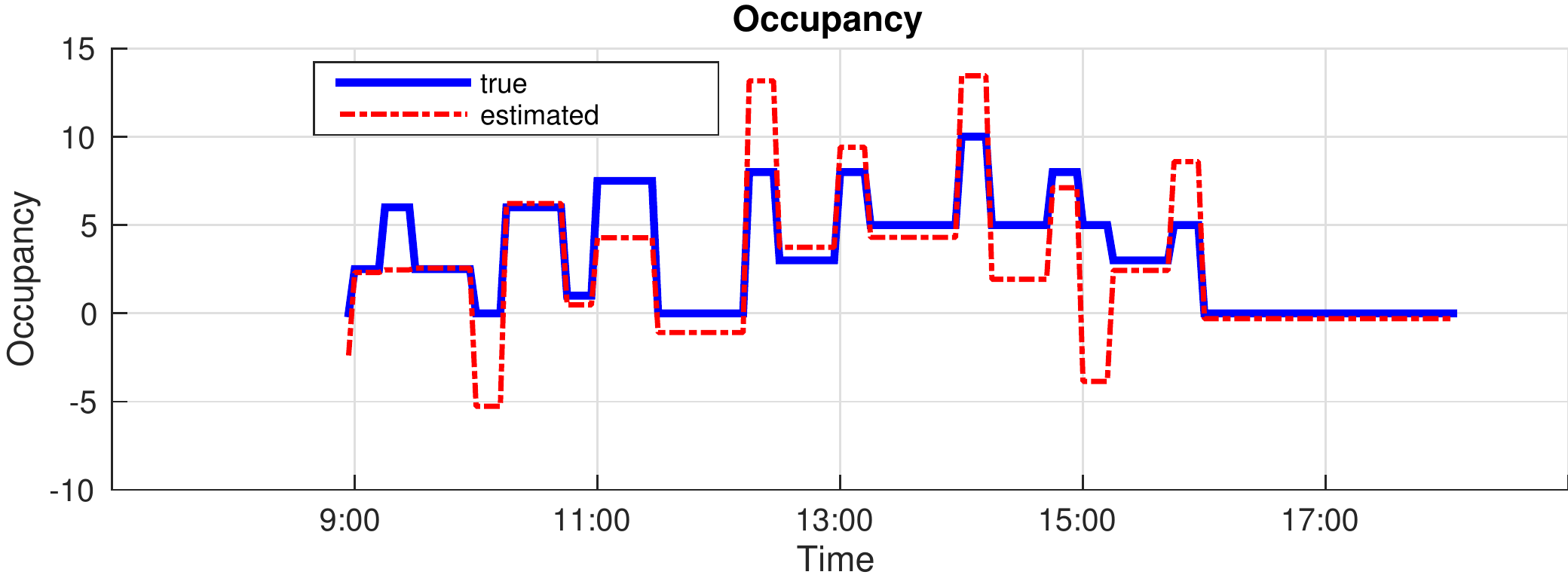}
\caption{True (noiseless) and predicted (from the identified model) CO$_2$ and occupancy signals. Left: database \texttt{kth\_mowc} (medium occupancy and closed windows), Tuesday. Right: database \texttt{kth\_mowo} (medium occupancy and open windows), Tuesday.}
\label{fig:room_CO2_concentration}
\end{figure*}
We evaluate the performance of the blind system identification algorithms described in the previous section on the data sets. Specifically, we run the identification algorithms using daily data records and discarding data before 9:00 and after 18:00. Also, we do not consider data regarding Saturday and Sunday, since the room is known to be empty. So, for each data set, we obtain 5 separate identification problems, one for each weekday. We define two accuracy scores.

\begin{enumerate}
\item The fit of the CO$_2$ signal, i.e.
\begin{equation}
  FIT_{CO_2} = 1 - \sqrt{\frac{\sum_{t=1}^{270} {(\widehat{CO}_2(t) -
  CO_2(t))}^2}{\sum_{t=1}^{270} {(CO_2(t) -  \mathrm{mean}[CO_2(t)])}^2}} \,,
\end{equation}
where $\widehat{CO}_2(t)$ is the output predicted by the identified model and 270 is the number of samples per interval considered.
\item The fit of the occupancy signal, namely
\begin{equation}
  FIT_{O} = 1 - \sqrt{\frac{\sum_{t=1}^{270} {(\widehat{O}(t) -
  O(t))}^2}{\sum_{t=1}^{270} {O(t)}^2}} \,.
\end{equation}
Note that the average value of the true occupancy is not removed in the denominator.
\end{enumerate}
The overparameterization method is not able to capture the CO$_2$ dynamics nor reconstruct the occupancy pattern, always giving negative fits. Thus we do not report its results.
The identification performance of the kernel-based method is summarized in
Table 2, where the average (over the days) daily fits are reported. Two
examples of the resulting outcomes are shown in Fig. 5, where we see the
CO$_2$ profile predicted by the identified model, compared with the noiseless
CO$_2$ profile in the dataset.
When windows are kept closed, the reconstruction performance is satisfactory, giving fits ranging from 89.72 \% to 98.44 \% in the CO$_2$, and fits ranging from 72.86 \% to 87.57 \% in the occupancy. However, when open windows are simulated, the fits drop to 80.3  $\div$ 87.24 \% in the CO$_2$ and 37.85  $\div$ 67.13 \% in the occupancy. This indicates that the effect of open windows cannot be neglected when trying to perform blind identification of the occupancy/CO$_2$ relation.
\begin{table}[h!]\label{tab:fits}
\begin{center}
\begin{tabular}{lcc}
  \toprule
Database & Average occupancy fit (\%)& Average CO$_2$ fit (\%) \\
\midrule
\texttt{kth\_lowc} & 87.6 & 98.4 \\
\texttt{kth\_mowc} & 76.2 & 92.8 \\
\texttt{kth\_howc} & 72.9 & 89.7 \\
\texttt{kth\_lowo} & 37.9 & 80.3 \\
\texttt{kth\_mowo} & 39.4 & 87.2 \\
\texttt{kth\_howo} & 67.1 & 86.0 \\
\bottomrule
\end{tabular}
\vspace{2mm}
\caption{Average occupancy and prediction fits for the different databases.}
\end{center}
\end{table}

\section{Discussion}
We have proposed and described a set of data generated from a simulated office
environment. The data set is targeted around those signals involved in the
CO$_2$ dynamics, such as ventilation inflow, window opening and number of
people in the room. Simulations were carried out using the commercial software
IDA-ICE and were shown to well-describe a real laboratory at KTH\@. We have
sketched a schematic representation of the environment, pointing out some
interesting problems from the system identification perspective. Among these
problems, we have attempted a blind identification of the dynamic relation
between the (unknown) number of occupants and the CO$_2$ signal.

We believe that the presented data can be potentially very interesting for the
system identification community, due to the numerous challenges arising from
this framework. This also holds true for researchers working in smart building
design, where the integration of smart devices with the building has made
data-based modeling techniques of paramount importance.

This data set is continuously evolving: we plan to perform further
simulations taking into account other aspects of the office environment, such
as external influences (solar radiation, outdoor temperature) and temperature
dynamics.

\section*{Appendix: Useful Parameters}

\begin{table}[h!]\label{tab:parameters}
\begin{center}
\begin{tabular}{llc}
\toprule
Parameter &  & Value \\
\midrule
Room height &[m] & 2.9\\
Floor area &[m$^2$] &  80\\
Door area &[m$^2$] & 1.6\\
Total window area & [m$^2$] & 2.56\\
Number of windows &[-] & 4\\
Minimum mechanical ventilation flow& [m$^3$/s] & 0.08 \\
Maximum mechanical ventilation flow& [m$^3$/s] & 0.28 \\
Building air tightness& [ACH $@$ 50 Pa] & 1.5 \\
Occupant activity level& [MET] & 1.8 \\
Maximum tuna fish weight  & [kg] & 684 \\
Outdoor air CO$_2$ concentration& [ppm] & 420\\
Sampling time of the data& [min] & 3\\
\bottomrule
\end{tabular}
\vspace{2mm}
\caption{Some parameters of interest for the room dynamics.}
\end{center}
\end{table}

\appendix

\printbibliography{}
 \end{document}